A BAYESIAN APPROACH TO POWER-SPECTRUM SIGNIFICANCE ESTIMATION, WITH APPLICATION TO SOLAR NEUTRINO DATA

P.A. STURROCK

Center for Space Science and Astrophysics, Varian 302,

Stanford University, Stanford, CA 94305-4060, U.S.A.

*(e-mail: sturrock@stanford.edu)*

**Abstract.** The usual procedure for estimating the significance of a peak in a power spectrum is to calculate the probability of obtaining that value or a larger value by chance, on the assumption that the time series contains only noise (e.g. that the measurements were derived from random samplings of a Gaussian distribution). However, it is known that one should regard this "P-Value" approach with caution. As an alternative, we here examine a Bayesian approach to estimating the significance of a peak in a power spectrum. This approach requires that we consider explicitly the hypothesis that the time series contains a periodic signal as well as noise. The challenge is to identify a probability distribution function for the power that is appropriate for this hypothesis. We propose what seem to be reasonable conditions to require of this function, and then propose a simple function that meets these requirements. We also propose a consistency condition, and check to see that our function satisfies this condition. We find that the Bayesian significance estimates are considerably more conservative than the conventional estimates. We apply this procedure to three recent analyses of solar neutrino data: (a) bimodality of GALLEX data; (b) power spectrum analysis of Super-Kamiokande data; and (c) the combined analysis of radiochemical neutrino data and irradiance data.

# 1. Introduction

The usual approach to significance estimation of power spectra is to compute the probability of obtaining the specified power or more on the basis of the null hypothesis that the time series consists only of noise. For the familiar assumption that the noise is Gaussian, the probability of obtaining power S in the range $S$ to $S + dS$ is given by

$$P_S(S|H0)dS = e^{-S}dS , \qquad (1)$$

and the probability of getting S or more is given by

$$P_>(S|H0) = \int_S^\infty dz\, P_S(z|H0) = e^{-S} . \qquad (2)$$

(See, for instance, Scargle 1982.) Conventional significance estimates are based on Equation (2).

The probability of getting a certain result or a "more extreme" result on the basis of a "null hypothesis" (as in Equation (2)) is known as the "P-Value." Textbooks on statistics (see, for instance, Utts, 1996) warn that this should not be interpreted as the probability that the null hypothesis is true.

However, one really wishes to know whether or not the null hypothesis is true. One can seek to obtain this information by using a Bayesian approach. (See, for instance, Good, 1983; Howson and Urbach, 1989; Jaynes, 2004; Sturrock, 1973, 1994.) The key point of this approach is that it requires us to specify a complete set of hypotheses, not just a single hypothesis. For instance, in analyzing the Bernoulli (coin-flipping) problem, one needs to consider not only the hypothesis that the coin is unbiased, but also the hypothesis that it is biased (Sturrock, 1997). In analyzing that problem, one finds that the Bayesian analysis leads to significance estimates that are much more conservative than those suggested by the P-Value estimates.

We set out the basic equations for a Bayesian analysis in Section 2. It should perhaps be emphasized that this analysis begins with an estimate of the power derived from power-spectrum analysis of a time series, not with a time series. As in our analysis of procedures for combining power estimates (Sturrock, Scargle, Walther, and Wheatland, 2005), we here assume that we are given the results of a power-spectrum analysis, but we have no access to the time series from which those results were derived. In Section 3, we set out the basic requirements for a Bayesian assessment of the power spectrum associated with a periodic signal, and propose a formula that meets these requirements. There is more than one way to combine assessments of two independent measurements of a power spectrum. We discuss two such procedures in Section 4, and check to see that, using the formula of Section 3, the results of these two procedures are reasonably consistent.

In power-spectrum analyses, one is usually examining the power over a range of frequencies, rather than at a single frequency. In Section 5, we therefore consider the significance to be assigned to the biggest peak in a band of frequencies. This procedure may be compared with the conventional procedure of computing the "false-alarm" frequency (Scargle, 1982). In Section 6, we apply our new procedure to a re-examination of (a) the bimodality of GALLEX measurements; (b) the power spectrum derived from Super-Kamiokande data; and (c) a combined analysis of Homestake and GALLEX radiochemical data and ACRIM irradiance data.

## 2. Basic Equations

In order to estimate the probability that a certain value of the power S is due to noise (i.e., that the null hypothesis is true), we need to add another hypothesis, H1, that the power S is not due to noise, i.e. that it is due to a periodic signal in the time series. If we can determine the probability distribution function $P_S(S|H1)$ for S, we may derive the probability that the time series contains a periodic signal by Bayes' theorem (Good, 1983; Howson and Urbach, 1989; Jaynes, 2004; Sturrock, 1973, 1994)

$$P(H1\,|\,S) = \frac{P_S(S\,|\,H1)}{P_S(S\,|\,H0)P(H0\,|-) + P_S(S\,|\,H1)P(H1\,|-)} P(H1\,|-), \qquad (3)$$

where $P(H0\,|-)$ and $P(H1\,|-)$ are the prior probabilities of H0 and H1.

In the absence of information that would lead one to favor H0 over H1 or vice versa, one is led to assign equal prior probabilities to H0 and H1, so that Equation (3) becomes

$$P(H1\,|\,S) = \frac{P_S(S\,|\,H1)}{P_S(S\,|\,H0) + P_S(S\,|\,H1)}. \qquad (4)$$

Similarly,

$$P(H0\,|\,S) = \frac{P_S(S\,|\,H0)}{P_S(S\,|\,H0) + P_S(S\,|\,H1)}. \qquad (5)$$

(Note that we now need consider only the actual value of the power; it is not necessary to consider the probability that the power is "S or more.")

We see that

$$\Omega(H0\,|\,S) \equiv \frac{P(H0\,|\,S)}{P(H1\,|\,S)} = \frac{P_S(S\,|\,H0)}{P_S(S\,|\,H1)}. \qquad (6)$$

Since H1 is the same as "not-H0," $\Omega(H0\,|\,S)$ is the "odds" on H0, based on the measurement S. In terms of the "log-odds" (Good, 1983) defined by

$$\Lambda(H0\,|\,S) = \log_{10}(\Omega(H0\,|\,S)), \qquad (7)$$

Equation (5) becomes

$$\Lambda(H0 \mid S) = \log_{10}\left(\frac{P_S(S \mid H0)}{P_S(S \mid H1)}\right). \tag{8}$$

The odds for H1 is the inverse of the odds or H0, and the log-odds for H1 is the negative of the log-odds for H0. We may if necessary always retrieve the probability from the odds:

$$P = \frac{\Omega}{\Omega + 1}. \tag{9}$$

## 3. Basic Requirements

We usually know the function $P_S(S \mid H0)$. For Gaussian noise, for instance, it is given by Equation (1). The challenge is to find an appropriate form of the function $P_S(S \mid H1)$. We now propose what seem to be reasonable requirements to put upon this function, and suggest a formula that seems to meet these requirements.

If we were given specific information about the possible a periodic signal (or signals) in the time series – for instance, the type of noise, the number of oscillations, and the possible amplitudes of the oscillations – it might be possible (but perhaps not easy) to estimate $P_S(S \mid H1)$ from that information. However, we here suppose that we have no detailed information about the possible oscillations. Then there is no obvious basis for calculating $P_S(S \mid H1)$. Does this mean that we can do nothing?

It seems reasonable to look for a functional form for $P_S(S \mid H1)$ that has the following properties:

a . It is nonzero for all values of S,
b . It is a monotonically decreasing function of S,
c . The rate of decrease is as slow as possible, and d . Its integral is unity.

Requirements (a) and (b) suggest that we adopt an inverse power law for P$_S$:

$$P_S(S|H1) = \frac{A}{(B+S)^\beta}, \qquad (10)$$

where $B > 0$. To meet requirement ( c ), we would like $\beta$ to be as small as possible, but we can meet requirement (d) only if $\beta > 1$.

If we adopt $\beta = 1$, the integral diverges, but only logarithmically. This suggests that we set an upper limit S$_M$ on the range of powers that we need to consider. Since we rarely encounter power spectra with $S > 20$, we propose adopting $\beta = 1$ and $S_M = 20$. Then A is determined by requirement (d), i.e.

$$A = \frac{1}{\ln(1 + S_M/B)} . \qquad (11)$$

The resulting log-odds of H0 is shown, as a function of S and for the choice $B = 1$, in Figure 1. We also show in this figure the values of the log-odds that we would find by adopting $S_M = 30$. We see that the difference between adopting $S_M = 20$ and $S_M = 30$ (the difference is only 0.05) is negligible. Hence making the choice $S_M = 20$ is in practice not restrictive and may be ignored.

## 4. Consistency Condition

Suppose that we have repeated an experiment or an observation, and so obtained two independent measurements of the power, $S_A$ and $S_B$, at a given frequency. From these two values, we may form the corresponding values of the log-odds: $\Lambda_A$ and $\Lambda_B$. Then the log-odds for the two power values, taken together, is given by

$$\Lambda_{S,AB} = \Lambda_A + \Lambda_B , \qquad (12)$$

where the subscript S indicates that this estimate is formed by summing the two log-odds values.

On the other hand, we may alternatively proceed by combining the two values of the power, and then forming the corresponding log-odds. We have shown elsewhere (Sturrock, Scargle, Walther, and Wheatland, 2005) that, from two power estimates, $S_A$ and $S_B$, we may form a statistic that is a function of the sum of the two powers, and is distributed exponentially. If we write

$$Z_{AB} = S_A + S_B , \qquad (13)$$

then this statistic (which we term the "Combined Power Statistic") is given by

$$G_{AB} = Z_{AB} - \ln(1 + Z_{AB}) . \qquad (14)$$

[This way of combining powers is equivalent to a procedure (see, for instance, Rosenthal, 1984) attributed to Fisher for combining P-Values.] We may use Equation (8) to form from $G_{AB}$ another estimate of the log-odds derived from the two power values,

$$\Lambda_{C,AB} = \log_{10}\left(\frac{P_S(G_{AB}|H0)}{P_S(G_{AB}|H1)}\right), \qquad (15)$$

where the subscript C indicates that this estimate is formed from the combined power statistic.

The ideal solution would be to find a functional form for $P_S(S|H1)$ which guarantees that $\Lambda_{S,AB} = \Lambda_{C,AB}$ for all values of $S_A$ and $S_B$. [This problem is left as an exercise for the reader!] We here simply check to see if our proposed form of $P_S(S|H1)$, given by Equation (11), is reasonably compatible with this consistency condition. This procedure implicitly also leads to consistency in combining three or more power spectra, since we may begin by combining just two power spectra and then combine the result with one more power spectrum, etc.

We show in Figure 2 the comparison of $\Lambda_C$ and $\Lambda_S$ for B = 1, for $S_1$ in the range 0 to 20, and for four values of $S_2$: 2, 4, 8 and 16. We see that, for most combinations of powers, the agreement is quite good. For $S_1, S_2 \geq 4$, the maximum discrepancy is only about 0.3.

We have also considered the range 0 to 20 for both $S_1$ and $S_2$, and we find that the maximum discrepancy is a minimum for $B = 1.65$, for which the discrepancy approaches 0.6 for small values of both $S_1$ and $S_2$. We show in Figure 3 the comparison of $\Lambda_C$ and $\Lambda_S$ for $B = 1.65$, for $S_1$ in the range 0 to 20, and for $S_2 = 2, 4, 8$ and 16. We find that, for $S_1, S_2 \geq 4$, the maximum discrepancy is only about 0.2.

Adopting $B = 1.65$, we find that, to sufficient accuracy, we may adopt $A = 0.4$ so that Equation (11) becomes, numerically,

$$P_S(S|H1) = \frac{0.4}{1.65 + S} \ . \tag{16}$$

Hence Equation (8) leads to the following formula for the log-odds of H0:

$$\Lambda(H0|S) = \log_{10}\left(2.5\,(1.65 + S)e^{-S}\right). \tag{17}$$

## 5. Significance of a Peak in a Range of Frequencies

In many applications, we are concerned not with evaluating the significance a peak at a single specified frequency, but with evaluating the significance of the biggest peak in a band of frequencies. The power spectrum of a time series of finite duration will have a finite number N of independent peaks. We therefore consider the following hypotheses:

H00: the principal peak is due to noise,
H11: the principal peak is due to a periodic signal.

On the basis of H00, since there are N possible frequencies, the probability of obtaining power in the range $S$ to $S + dS$ at one of the frequencies is

$$P_S(S|H00)dS = NP_S(S|H0)dS \ . \tag{18}$$

On the basis of H11, the probability that we find power in the range $S\ to\ S+dS$ due to the fact that the principal peak is due to a periodic signal is

$$P_S(S|H11)dS = P_S(S|H1)dS . \tag{19}$$

Hence the log-odds of H00 is given by

$$\Lambda(H00|N,S) = \log_{10}\left(\frac{N(B+S)}{Ae^S}\right) \tag{20}$$

or, numerically,

$$\Lambda(H00|N,S) = \log_{10}\left(2.5N(1.65+S)e^{-S}\right). \tag{21}$$

## 6. Examples and Discussion

We now return to a comparison of the usual significance estimate in terms of a P-Value with the significance estimate obtained by our Bayesian approach. We show in Figure 4 the log-odds of H0 (the assumption that there is no periodic signal) as a function of the power S, for the choice $\beta=1$, $B=1.65$, and $S_M=20$. We show in the same figure the logarithm of $e^{-S}$, which is seen to be smaller than the log-odds of H0 by a factor between 4 and 60, with a median value of about 30. The same data are given in Table 1. It is clear that the usual "P-Value" significance estimation should not be confused with an estimate of the probability that the time-series contains no periodic signal.

It is interesting to note that $\Lambda(H0|S) = \Lambda(H1|S) = 0$ for $S=2.33$. Hence we need $S>2.33$ for a measurement to begin to favor a periodic signal. We also see that a P-Value of 0.05 corresponds to a power $S=3.00$, for which the log-odds for H0 is only 0.21, for an odds of 1.58 ($P=0.61$) giving no evidence in favor of a periodic signal. In order for the odds estimate to favor a periodic signal by 20:1, we need $S=7.83$, for which the corresponding P-Value estimate is 0.0004.

We see that Bayesian significance estimates are much more conservative than the conventional P-Value estimates.

Prior probabilities play no role in the usual P-value approach to significance estimation, but they are crucial for a Bayesian estimation. In Section 2, we began by proposing that we adopt equal prior probabilities for H0 and H1. This seems to be the appropriate choice if one has no prior information concerning these two hypotheses. However, there may be prior information that should be taken into account. One way to do this is to adopt different values for $P(H0|-)$ and $P(H1|-)$. One could attempt to modify equations accordingly, but this approach would require reconsideration of the consistency condition for each new choice of the prior probabilities, which would be a distraction in data analysis.

There is an alternative and more convenient procedure for taking account of prior information (Sturrock, 1994). We can begin by saying that, in the absence of relevant prior information, it is appropriate to assign equal probabilities to H0 and H1. (This is the "maximum entropy" procedure.) However, if relevant "advance" information A then becomes available, we can analyze this information explicitly and calculate the corresponding probabilities $P(H0|A)$ and $P(H1|A)$. From these probabilities, we may calculate $\Omega(H0|A)$, the odds on H0, from

$$\Omega(H0|A) = \frac{P(H0|A)}{P(H1|A)}, \tag{22}$$

and the log-odds corresponding to the advance information from

$$\Lambda(H0|A) = \log_{10}(\Omega(H0|A)) = \log_{10}(P(H0|A)/P(H1|A)). \tag{23}$$

We may now take account of the advance information simply by adding the log-odds given by Equation (23) to the log-odds given by Equation (8):

$$\Lambda(H0|A,S) = \Lambda(H0|A) + \Lambda(H0|S). \tag{24}$$

With this approach, there is no need to revisit the consistency condition. We can simply retain the analysis and results of Section 4.

Consider, as an example, $S = 10$. We see from Table 1 that $\log_{10}(P) = -4.34$, whereas our Bayesian analysis gives $\Lambda(H0|S) = -2.87$. If we were to begin by assuming that the advance information corresponds to an odds of 10:1 in favor of H0 (corresponding to $P(H0|A) = 0.91$ and $P(H1|A) = 0.09$), then we see from Equations (22) and (23) that $\Lambda(H0|A) = 1$ so that, from Equation (24), the log odds for H0, based on the power value and the advance information, becomes $\Lambda(H0|S,A) = -1.87$. Conversely, if we were to begin by assuming that the advance information corresponds to an odds of 10:1 in favor of H1, we would find that $\Lambda(H0|S,A) = -3.87$. It should not be surprising that significant advance information has a significant effect on the end result.

It is interesting to reverse the argument and ask what prior information would be required to reconcile the P-Value estimate with the Bayesian estimate. We would then set $\Lambda(H0|S,A) = -4.34$. Since $\Lambda(H0|S) = -2.87$, we see from Equation (24) that $\Lambda(H0|A) = 1.47$, so that $\Omega(H0|A) = 29.5$. advance information in favor of H0 for the P-value approach to give the same estimate as we would obtain from our Bayesian approach.

As an example of the application of our formulae to power estimation at a single frequency, we consider our recent analysis of bimodality of GALLEX (Anselmann *et al.*, 1993; Anselmann *et al.*, 1995; Hampel *et al.*, 1996; Hampel *et al.*, 1999) and GNO (Altmann *et al.*, 2000; Kirsten *et al.*, 2003; Altmann *et al.*, 2005) solar neutrino data (Sturrock, 2008a). This analysis is in effect a search for a specific periodicity in a histogram formed from capture rate data, normalized to conform to a normal distribution. Our analysis led to the estimate $Bim = 8.50$ for the bimodality index. According to the conventional P-Value approach, we conclude that there is probability 0.0002 of obtaining this result from normally distributed random numbers. However, we see from Equation (17) that the odds on the null (no bimodality) hypothesis is 0.005. Hence the Bayesian analysis yields the result that we have established the case for bimodality with confidence 0.995, i.e. at the 99.5% confidence level. This is an interesting result, but more conservative than the confidence level 0.9998 (99.98%) suggested by the conventional P-Value approach.

As a second example, we consider the application of the Bayesian approach to the power spectrum derived from Super-Kamiokande (Fukuda et al., 2001, 2002; Fukuda, 2003) solar neutrino data (Sturrock, Caldwell, Scargle, and Wheatland, 2005). We found that there was a notable peak (at frequency $9.43\,yr^{-1}$) with power $S = 11.67$. Examination of the power spectrum over the search band $1-36\,yr^{-1}$ (the widest band compatible with no aliasing) shows that the number of independent peaks has the value $N = 126$. On inserting these figures into Equation (21), we obtain the result $\Lambda(H00\,|\,N,S) = -1.45$, corresponding to an odds ratio of 0.036. Hence we obtain the estimate that the probability that there is no periodic signal is 0.04.

We may compare this with the usual "false alarm" probability estimate (Scargle, 1982), given by

$$P(\text{false alarm}) = 1 - \left(1 - e^{-S}\right)^N . \qquad (25)$$

This corresponds to $P(H0\,|\,A) = 0.966$ and $P(H1\,|\,A) = 0.034$. We would have to have significant

We find that the false alarm probability is 0.0011. Once again, we find the Bayesian significance estimate to be considerably more conservative than the conventional estimate.

As a third example, we consider the recent combined analysis (Sturrock 2008b) of Homestake (Davis, 1996; Davis, Harmer, and Hoffman, 1968; Cleveland et al., 1998) and GALLEX (Anselmann *et al*., 1993, 1995; Hampel *et al.,* 1996, 1999) radiochemical data and ACRIM irradiance data (Willson 1979, 2001; www.acrim.com)). We have formed the joint power statistic (JPS; Sturrock, Scargle, Walther, and Wheatland, 2005) from four independent datasets: the Homestake neutrino data, the GALLEX neutrino data, ACRIM data for the Homestake time interval, and ACRIM data for the GALLEX time interval. For four power spectra, the joint power statistic is given - to sufficient accuracy - by

$$J = \frac{3.88 Y^2}{1.27 + Y}, \qquad (26)$$

where

$$Y = (S_1 * S_2 * S_3 * S_4)^{1/4}. \qquad (27)$$

We find a striking peak (J = 40.87) in the JPS spectrum at 11.85 year$^{-1}$. We interpret this frequency as the synodic rotation frequency of the core, corresponding to a sidereal rotation frequency of 12.85 year$^{-1}$, or 407 nHz. Since the JPS is designed to have the same exponential distribution as the individual powers from which it is formed, we may evaluate its significance by the conventional procedure and by our Bayesian procedure.

Adopting a search band of 10 – 15 year$^{-1}$ for rotational frequencies, we find that there are 66 peaks in the JPS in that band. The false-alarm formula of Equation (22) then leads to the estimate of 1.2 10$^{-16}$ for obtaining the value J = 40.87 or higher by chance. On the other hand, Equation (21) leads to a log-odds value of -13.90, corresponding to an odds value of 1.25 10$^{-14}$ for hypothesis H00. (The probability has essentially the same value.) This is more conservative than the false-alarm frequency estimate, but it still represents very strong evidence that the neutrino and irradiance data are subject to a common periodic signal.

No doubt, this is not the final answer to the challenge of devising a Bayesian procedure for the significance estimation of power spectra. However, this initial attempt may help point the way to something better.

## Acknowledgements

Thanks are due to Alexander Kosovichev, Jeffrey Scargle and Guenther Walther for helpful discussions related to this work, which was supported by NSF Grant AST-0607572.

FIGURES

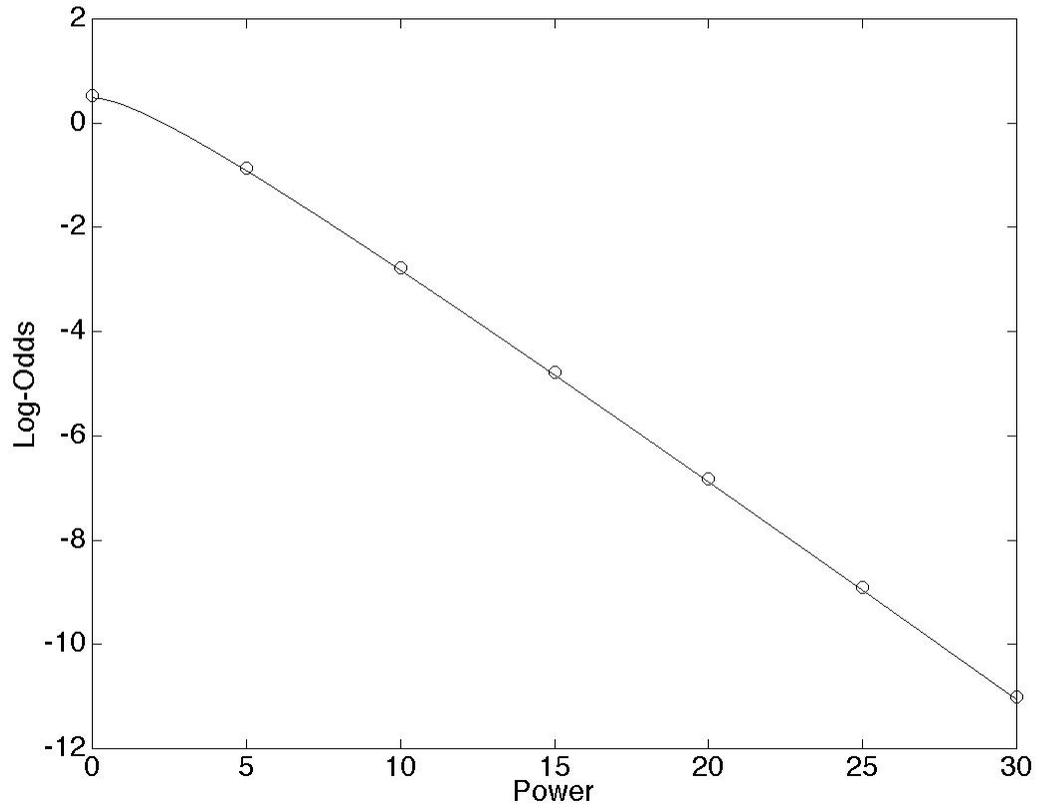

Figure 1. The log-odds of H0 as a function of S for the choice $\beta = 1$, $B = 1$ and $S_M = 20$. The figure also shows, as circles, the values of the log-odds computed with $S_M = 30$.

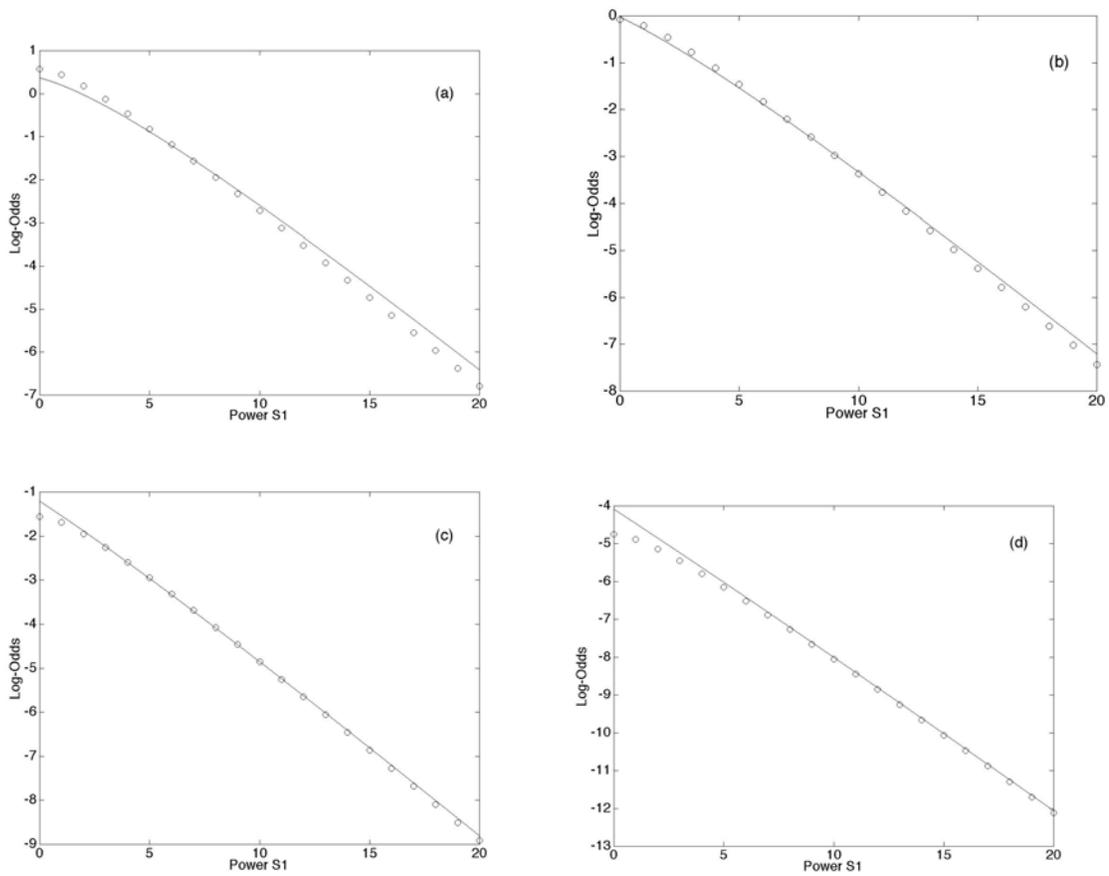

Figure 2. The log-odds, for H0, of the combined power statistic formed from S1 (the abscissa) and S2, where $B = 1$ and (a) $S2 = 2$, (b) $S2 = 4$, (c) $S2 = 8$, (d) $S2 = 16$. The sum of the log-odds for H0, calculated separately from S1 and S2, is shown as circles.

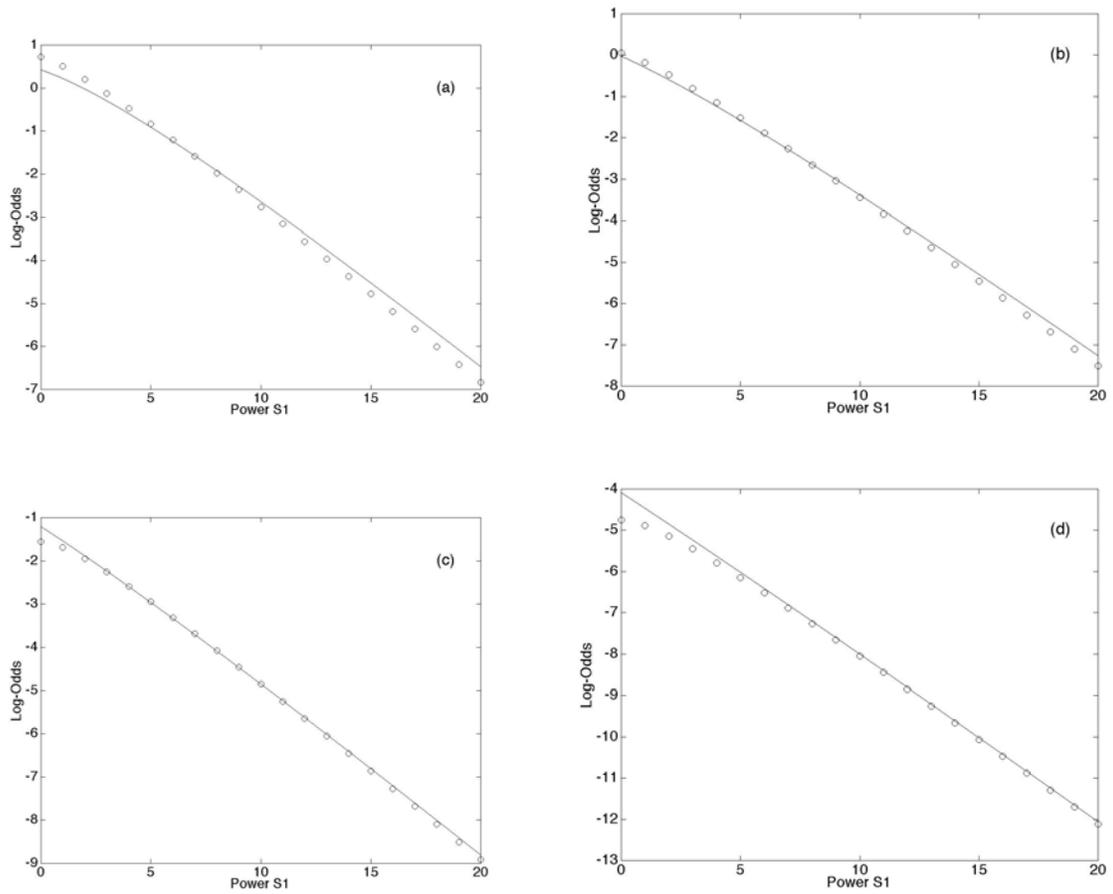

Figure 3. The log-odds, for H0, of the combined power statistic formed from S1 (the abscissa) and S2, where $B = 1.65$ and (a) S2 = 2, (b) S2 = 4, (c) S2 = 8, (d) S2 = 16. The sum of the log-odds for H0, calculated separately from S1 and S2, is shown as circles.

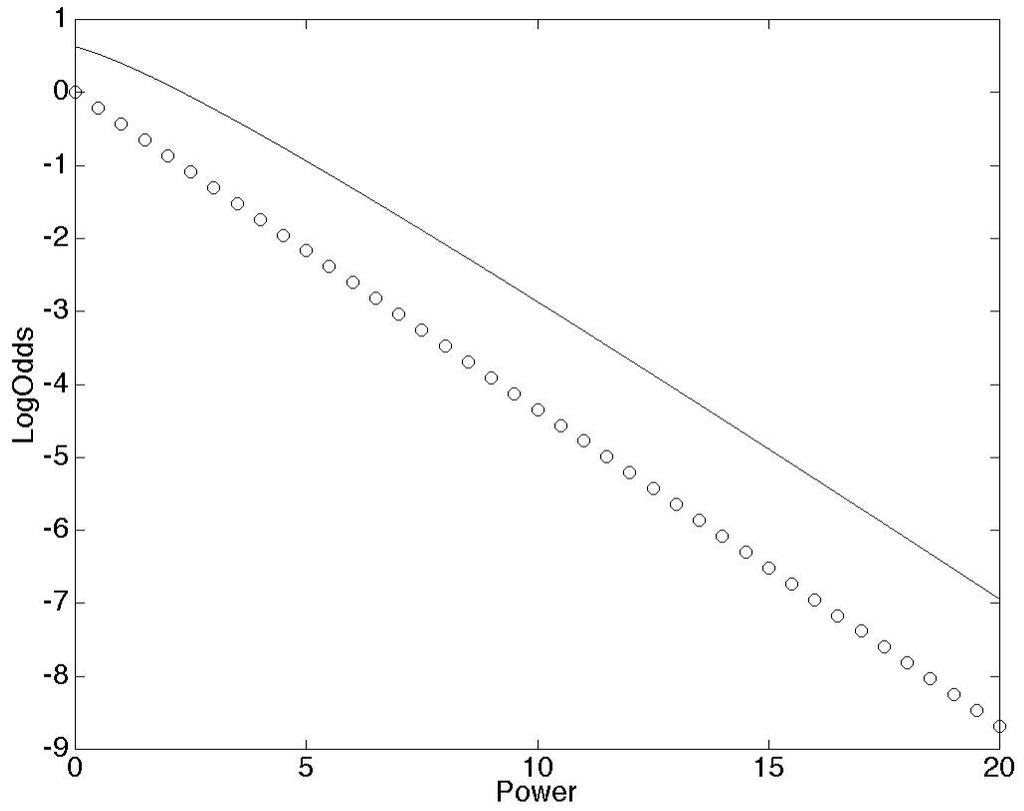

Figure 4. The log-odds of H0 as a function of S for the choice $\beta = 1$, $B = 1.65$, and $S_M = 20$. The figure also shows, as circles, the logarithm of .

TABLE

Table 1. Comparison of $\text{Log}10(e^{-S})$ and the LogOdds of the Null Hypothesis

| S | $\text{Log}10(e^{-S})$ | Log Odds |
|---|---|---|
| 0 | 0.00 | 0.63 |
| 1 | -0.43 | 0.40 |
| 2 | -0.87 | 0.10 |
| 3 | -1.30 | -0.23 |
| 4 | -1.74 | -0.58 |
| 5 | -2.17 | -0.94 |
| 6 | -2.61 | -1.31 |
| 7 | -3.04 | -1.69 |
| 8 | -3.47 | -2.08 |
| 9 | -3.91 | -2.47 |
| 10 | -4.34 | -2.87 |
| 11 | -4.78 | -3.27 |
| 12 | -5.21 | -3.67 |
| 13 | -5.65 | -4.07 |
| 14 | -6.08 | -4.48 |
| 15 | -6.51 | -4.88 |
| 16 | -6.95 | -5.29 |
| 17 | -7.38 | -5.70 |
| 18 | -7.82 | -6.12 |
| 20 | -8.69 | -6.94 |